%% file: main.tex
\documentclass[a4paper]{article}

\usepackage{adjustbox}

\usepackage[utf8]{inputenc}
\usepackage{makecell}

\usepackage{algorithm}
\usepackage{algorithmic}

\usepackage{caption}
\usepackage{subcaption}

\usepackage{tikz}
\usetikzlibrary{shapes}

\usepackage{bbold}
\usepackage{import}
\usepackage{hyperref}

\usepackage{enumitem}
\setlist[itemize]{leftmargin=*}

\usepackage{array}

\makeatletter
\newcommand{\thickhline}{%
    \noalign {\ifnum 0=`}\fi \hrule height 1pt
    \futurelet \reserved@a \@xhline
}
\newcolumntype{"}{@{\hskip\tabcolsep\vrule width 1pt\hskip\tabcolsep}}
\makeatother

\title{Diverse Reviewer Suggestion for Extending Conference Program Committees}

\author{ \href{https://orcid.org/0000-0002-5075-7699}{\includegraphics[scale=0.06]{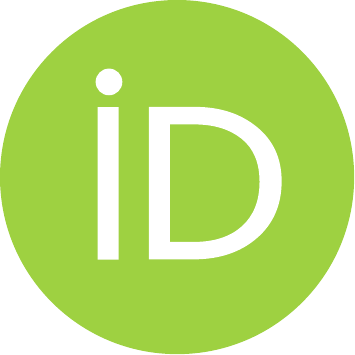}\hspace{1mm}Christin Katharina Kreutz$^{1}$}, \href{https://orcid.org/0000-0003-2762-721X}{\includegraphics[scale=0.06]{orcid.pdf}\hspace{1mm}Krisztian Balog$^{2}$}, \href{https://orcid.org/0000-0001-5379-5191}{\includegraphics[scale=0.06]{orcid.pdf}\hspace{1mm}Ralf Schenkel$^{3}$} \\
	$^{1}$ \texttt{kreutzch@uni-trier.de}, Trier University, Germany\\
	$^{2}$ \texttt{krisztian.balog@uis.no}, University of Stavanger, Norway \\
	$^{3}$ \texttt{schenkel@uni-trier.de}, Trier University, Germany\\
}

\date{}

\providecommand{\keywords}[1]
{
	\small	
	\textbf{\textit{Keywords---}} #1
}

\begin{document}
\maketitle

\begin{abstract}
Automated reviewer recommendation for scientific conferences currently relies on the assumption that the program committee has the necessary expertise to handle all submissions.  However, topical discrepancies between received submissions and reviewer candidates might lead to unreliable reviews or overburdening of reviewers, and may result in the rejection of high-quality papers.
In this work, we present DiveRS, an explainable flow-based reviewer assignment approach, which automatically generates reviewer assignments as well as suggestions for extending the current program committee with new reviewer candidates.
Our algorithm focuses on the diversity of the set of reviewers assigned to papers, which has been mostly disregarded in prior work.  Specifically, we consider diversity in terms of professional background, location and seniority. 
Using two real world conference datasets for evaluation, we show that DiveRS improves diversity compared to both real assignments and a state-of-the-art flow-based reviewer assignment approach.  Further, based on human assessments by former PC chairs, we find that DiveRS can effectively trade off some of the topical suitability in order to  construct more diverse reviewer assignments.
\end{abstract}

% keywords can be removed

\import{./}{1.Introduction.tex}

\import{./}{2.RelatedWork.tex}

\import{./}{3.ProblemSetting.tex}

\import{./}{4.Method.tex}

\import{./}{5.ExperimentalSetup.tex}

\import{./}{6.Experiments.tex}

\import{./}{7.Conclusion.tex}

\section*{Acknowledgements}
We want to thank the conference general chairs of ICTIR'19 and ICTIR'20 
%CIKM'21, ECIR'21, ICTIR'19 and ICTIR'20 
for kindly providing us with the conference data, as well as the PC chairs for evaluating our automatic recommendations.

\end{document}

%% file: 1.Introduction.tex
\section{Introduction}

% general setting
Scientific publishing heavily relies on peer review, which is typically performed by members of the program committee (PC) of a conference. In general, PCs need to grow and change each year: to keep up with the increasing number of submissions~\cite{serebrenik}, 
 to avoid tunnel vision as well as unchanging perceptions of good or bad concepts~\cite{DBLP:conf/nsdi/Douglis08}
(e.g., the ACM SIGSOFT policy recommends to change one third of the members each year~\cite{serebrenik}), and former PC members might become unavailable~\cite{DBLP:conf/cikm/HanJYH13}.
According to current practice, organisers compose the PC before the submission period of manuscripts ends. Once submissions are closed, %manuscripts are assigned to PC members for review~\cite{SBES}. 
each manuscript gets a number of PC members, also called reviewers, assigned by the PC chairs, either manually or automatically (based on bidding information or preferred topics, entered by reviewers)~\cite{DBLP:journals/corr/abs-1806-06237,tang,robustModel}.
%Reviewers oftentimes bid on their preferred topics or papers before the PC chairs conduct a final matching manually or automatically. 
Importantly, \textcolor{black}{to the best of our knowledge,} current approaches to reviewer recommendation assume a perfectly composed PC, and do not consider modification or extension as a necessity.

% problem
To the best of our knowledge, there is currently no way of reliably estimating the topical composition or amount of incoming submissions. Therefore, a previously disregarded problem is the possible mismatch between the expertise of current PC members and the expertise required for the assessment of all submissions.  This problem may be further amplified by the ever-changing PC.
%The always changing and expanding PCs may further amplify this problem lack of planning reliability.
%
% effects of problem
Consequences of the mismatch might result in manuscripts tackling topics far from the PC's interests being less favourably reviewed~\cite{SBES} and a general overburdening of reviewers. This, in turn, might lead to innovative and complex submissions being rejected solely due to low-quality reviews~\cite{birman} or failure to find errors in submissions~\cite{Severin2021.01.14.426539}. 

% solution + requirements
A solution for the above issues would be the inclusion of new and additional PC members after the submission period ended, but before the review assignments have been made.
This can especially help to cover new or emerging research topics~\cite{DBLP:conf/cikm/HanJYH13}
and to ensure that under-represented groups can gain exposure and reviewing experience~\cite{DBLP:journals/ccr/Sekar16}.
Identification of appropriate candidates is challenging as PC members should be diverse in localities, seniority~\cite{Severin2021.01.14.426539,SBES}, research topics and gender~\cite{SBES}.
%It would be desirable to appoint new members automatically.
%The appointment of new members should be conducted automatically, as manuscripts need to be reviewed rapidly~\cite{DBLP:conf/fun/BenderMS0V16}.
Furthermore, suggested candidates should be explainable, in order to aid the conference chairs in effective and efficient decision making.

% describe why this is a recommendation problem
%The reviewer coverage problem is a recommendation instead of a retrieval problem as there is no fixed information need, the underlying primarily textual data is semi-structured and the task is repeated periodically~\cite{DBLP:journals/cacm/BelkinC92}. 
In this paper we focus not only on the automatic assignments of reviewers to submissions (i.e., \emph{reviewer assignment}), but also introduce and address the problem of \emph{reviewer coverage}: ensuring the assignment of suitable reviewers to all submissions.  This gives rise to the novel task of \emph{reviewer suggestion} for PC extension: given the current PC and all submitted manuscripts of a venue, recommend new reviewer candidates to be added to the PC.
%
%novel task of suggesting additional PC members for inclusion, based on the set of submissions.
%a superordinate recommendation step: given a PC and a set of submissions, additional PC members need to be suggested for inclusion in the PC when a mismatch between those two is encountered.
%The overall goal of reviewer recommendation systems is to assign suitable reviewer sets for all submissions of the conference. To achieve this, here another superordinate recommendation step is encountered: given a PC and a set of submissions, additional PC members need to be identified when a mismatch between those two is encountered. 
Note that these two tasks are interconnected: 
our reviewer assignment method identifies submissions that would not receive adequate reviewers using the current PC, which in turn triggers the suggestion of new reviewers to extend the PC.  Those newly included persons should not only be capable of ideally assessing multiple manuscripts but also ensure diversity of the whole PC.
%
% identify gaps independent by the specific papers,
% and also for specific papers (based on reviewer assignment)
% research question, goal + contribution
Note that some gaps in the PC can be identified without requiring paper-reviewer assignments (e.g., not enough senior reviewers, reviewers from a given location or stark imbalance in academic vs. non-academic backgrounds of reviewers), while other gaps may only be identified once a (preliminary) assignment is done.

%To tackle this research question we define and study the \textbf{Reviewer Coverage Problem}: given a PC and all submitted manuscripts of a venue, the task is to recommend new reviewer candidates enabling a suitable assignment of reviewer sets for all submissions as well as creating a balanced, de-biased PC. 
%The suggestion of new PC members should be explainable to make them more trustworthy and engaging~\cite{DBLP:conf/recsys/McInerneyLHHBGM18}.

% general approach and design/definition decisions
The main contribution of this work is a flow-based reviewer suggestion and PC extension approach, termed \emph{DiveRS}.  
The main idea behind DiveRS is to iteratively identify submissions that are unlikely to get a set of suitable reviewers assigned.  These problematic submissions and currently underrepresented diversity aspects (professional background, location or seniority) determine the reviewer candidates for inclusion in the PC to support a feasible reviewer assignment.  We capture these characteristics in a constrained optimisation problem.
%Currently we do not treat the PC extension part of RCP as a learning problem due to data sparsity. The PC of previous conferences is mostly passed on to further instances without much additional information except for maybe the explicit exclusion of few candidates. Bidding information of PC members for submissions is not publicly visible in general. Thus, user interactions and ratings are not available for model training purposes.
At its core, DiveRS relies on a reviewer assignment method, which considers reviewers as a set for each paper, in order to satisfy diversity constraints.  Additionally, reviewers' individual upper and bounds of the numbers of papers to review, and their conflicts of interests, also need to be respected. 

%Our reviewing model can be described as follows: each submission is assigned a set of reviewers without conflicts of interest which are most similar to the submission and satisfy the following constraints: (1) Reviewers in the set are independent from each other and (2) contain at least one reviewer from a non-academia (industry) and one from an academia background; (3) Their location (in form of continent) cannot be all the same and (4) at least one senior researcher needs to be in every reviewer set; (5) Reviewers' individual upper and lower bounds of numbers of papers they can review are respected. 
% evaluation

We evaluate DiveRS on real-world conference datasets in two parts.  First, we compare it on the task of \emph{reviewer assignment} against the current state-of-the-art,  PR4All~\cite{DBLP:journals/corr/abs-1806-06237}, and against real assignments, in terms of both established measures (mean number of papers assigned, fairness, and textual diversity of reviewer sets) as well as novel measures (diversity and dependency between reviewers).
We show that DiveRS achieves fairness that is on par with PR4All, while being superior in terms of diversity.
Second, we evaluate the \emph{reviewer suggestion} task by asking actual PC chairs to assess the generated suggestions for PC extension in terms of relevance, usefulness, and accompanying explanation.  Our results indicate that DiveRS can effectively trade off topical suitability in order to improve the diversity of the assigned reviewer sets.

In summary, we make the following contributions:
\begin{itemize}
    \item We propose the reviewer coverage problem as an extension of the reviewer assignment problem, where we no longer assume the current PC to be perfectly suitable for all submissions.  We define the extension of the PC, to accommodate possibly ill-covered submissions, as part of the objective.
    \item We present DiveRS\footnote{\textcolor{black}{DiveRS implementation: \url{https://github.com/kreutzch/DiveRS}}}, a novel reviewer assignment and PC extension approach. It incorporates previously overlooked diversity aspects in terms of professional background, location and seniority of reviewer candidates directly in the assignment process, and generates explainable suggestions for extending the PC.
    %To the best of our knowledge our approach is the first one to identify likely problematic submissions which require for the introduction of new reviewers into the program committee. This enables the computation of a reviewer-submission assignment which is not only feasible but also of high diversity and fairness.
    %The newly suggested reviewer candidates are explainable.
    \item We propose new measures for evaluating the diversity and dependency of reviewer sets.
    \item We automatically evaluate our approach on two real-world datasets and demonstrate its suitability in manual evaluations with the actual PC chairs of these conferences.
\end{itemize}
%
%The implementation of DiveRS will be made publicly available upon acceptance.

% structure
%The paper is structured as follows: Section~\ref{rw} contains related work on reviewer assignment and program committees. In Section~\ref{problemStatement} we describe the observed problem and notation in detail before Section~\ref{method} tackles our proposed solution. Section~\ref{datasets} introduces our newly constructed datasets. The evaluation of our approach is described in Section~\ref{eval}.

%% file: 2.RelatedWork.tex
\section{Related Work}
\label{rw}

Areas related to our work are \emph{reviewer assignment}, which corresponds to the typical reviewer assignment problem, as well as the general field of \textit{program committee construction}, which relates to the extension of PCs. 
For conference organisers there are many systems supporting the bidding and reviewer assignment process but ``[e]xtending PCs based on submitted papers'' as identified as a future objective by Price and Flach~\cite{DBLP:journals/cacm/PriceF17} has not yet been tackled to the best of our knowledge.
There have been efforts to expand expert sets to hold more persons similar to the ones already contained in the set~\cite{DBLP:conf/ercimdl/VergoulisCDT20} but these approaches differ from our research objective: instead of finding more similar experts, our goal is to suggest an unbiased and diverse set of reviewer candidates to better cover the topical composition of incoming submissions.

%\subsection{Reviewer Assignment}
\textbf{Reviewer Assignment.}
There is a multiplicity of author-topic models to capture topical relationships between authors and (their) papers~\cite{DBLP:journals/jis/JinGMC19,DBLP:conf/sigir/Kawamae10a,DBLP:conf/airs/MouGJC15,DBLP:conf/uai/Rosen-ZviGSS04,DBLP:conf/coling/TuJRH10}. % which are often used for expert search. 
We refrain from discussing them in detail or utilising them here, as our focus within assigning reviewers to submissions lies not only on topical similarity of the two, but more on diversity aspects.

Conry et al.~\cite{DBLP:conf/recsys/ConryKR09} tackle the reviewer assignment problem with given bidding information as an optimisation problem with global criteria. They extend bidding data by predicting new preferences of reviewers, and utilise manuscript as well as reviewer similarities.
Liu et al.~\cite{robustModel} recommend $n$ reviewers for each manuscript which are dependent on each other. They model reviewers' expertise, authority and diversity in a graph, which they traverse with random walk with restart. The number of co-authorships is modelled as authority.
Tang et al.~\cite{tang} propose a constraint-based optimisation framework that proposes sets of reviewers for query manuscripts and user feedback, if available. They incorporate expertise matching, authority aspects based on seniority, load balance and aim to maximise the topic coverage between reviewer sets and manuscripts, using LDA. % For this they utilise LDA. Their definition of authority includes different expertise levels.
%
%Baskin and Krishnamurthi~\cite{DBLP:conf/recsys/BaskinK09} provide a solution for the aggregation of reviewer scores to determine an ordering of submitted manuscripts with the goal to help identifying the best or accepted ones.
%
Long et al.~\cite{DBLP:conf/icdm/LongWPY13} study topic coverage and fairness of manuscript-reviewer assignments. They maximise the numbers of different topics of manuscripts in which the assigned reviewer set is knowledgeable. Additionally, they define and regard the influence of different conflict of interest types, such as the competitor relationship, in the assignment.
Kou et al.~\cite{DBLP:conf/sigmod/KouUMG15} build upon~\cite{DBLP:conf/icdm/LongWPY13} and instead observe a weighted topic coverage score. % for sets of reviewers and the topics in manuscripts to quantify the quality of assignments. 
Their approach calculates the assignment resulting in the approximate maximum weight-coverage group-based scores, while fulfilling workload and reviewer set size constraints.

Jecmen et al.~\cite{DBLP:conf/nips/JecmenZLSCF20} provide a solution for the reviewer assignment problem, which focuses on supporting the integrity of the peer review process. The approach prevents reviewers' manipulation efforts in the assignment to either submit overly positive or negative feedback as well as de-anonymise the reviewing process. Here, the similarity between manuscripts and reviewers' profiles (expertise) is a critical factor in the randomised assignment.
Kobren et al.~\cite{DBLP:conf/kdd/KobrenSM19} introduce a paper-reviewer-assignment strategy which incorporates upper and lower load bounds per reviewer, guarantees a minimal required expertise in the area of the submission from all assigned reviewers and optimises a global objective. They present a linear programming and %as well as a 
min-cost flow-based heuristic approach.

The Toronto Paper Matching System (TPMS)~\cite{toronto} conducts automatic reviewer assignment for all manuscripts submitted to a conference by using either word count representation or LDA topics, but can also incorporate reviewers' bids on submissions. TPMS supports some constraints: papers must be reviewed by three reviewers, and reviewers are assigned not more than a certain limit of papers. Reviewers for manuscripts are determined based on expertise extracted from their published papers and maximising the similarity between reviewers and manuscripts.
Stelmakh et al.~\cite{DBLP:journals/corr/abs-1806-06237} use TPMS in PR4All; they propose an approach utilising a max-flow algorithm to identify the top-$k$ papers submitted to conferences, which should be accepted. They focus on fairly assigning suitable reviewer sets to all submissions via TPMS, especially those which received low similarity with all reviewer candidates. This approach is considered as the state of the art for flow-based reviewer assignment~\cite{DBLP:conf/kdd/KobrenSM19}.

We note that the datasets used in related work are mostly not available online and even fewer contain all submissions of a conference, i.e., include rejected papers.  
Those that remain either do not contain the real reviewers (ICLR 2018~\cite{DBLP:conf/nips/JecmenZLSCF20,DBLP:journals/corr/abs-1806-06237}) or do not contain names of both reviewers and authors (MIDL, CVPR and CVPR2018~\cite{DBLP:conf/kdd/KobrenSM19}).
Thus, to the best of our knowledge, there is no publicly available dataset including rejected papers, and non-anonymised reviewer and author names from a real conference.  
Therefore, we create our own datasets based on real conference data in \S\ref{sec:experimentalSetup:datasets}.

%No single dataset fulfills the four required properties to convincingly evaluate a reviewer assignment approach: publicly available dataset, inclusion of rejected papers, depiction of real conference data and inclusion of non-anonymised reviewer and author names.

%\subsection{Program Committee Construction}
\textbf{Program Committee Construction.}
Han et al.~\cite{DBLP:conf/cikm/HanJYH13} recommend PC members for conferences based on the previous year's PC and core authors, preferring candidates socially close to current chairs. They build a language model for a conference by aggregating previously published papers and compare it to PC candidates' publications. 
%PC candidates' publications are compared to the conference' language model, built by aggregating previously published papers. 
Authoritativeness of candidates influences the recommendations.
Sekar~\cite{DBLP:journals/ccr/Sekar16} introduces EZ-PC, a tool to define constraining factors and help automate the PC formation process as an integer linear programming problem.
Several factors are considered:
%in PC formulation: 
topical coverage, diversity of the PC, avoiding over-representation of groups and keeping the PC size manageable. 
\textcolor{black}{The main differences between their work and ours are that diversity constraints in EZ-PC are on the PC level, and they do not support reviewer assignment.}

% evaluation
%Avin et al.~\cite{DBLP:journals/snam/AvinLPT16} analyse co-authorship networks to search for bias of PCs in favour of collaborators. They observe two quality aspects of PCs: coverage (percentage of authors/papers connected to PC) and bias of accepting papers towards PC members' collaborators.

%Zhuang et al.~\cite{DBLP:conf/jcdl/ZhuangELG07} introduce conference quality measures based on PC members such as betweenness centrality or closeness centrality.

%Price and Flach~\cite{DBLP:journals/cacm/PriceF17} suggest construction of a recommendation system which expands current PCs of venues based on the submitted manuscripts.

%% file: 3.ProblemSetting.tex
\section{Problem Setting}
\label{problemStatement}

\subsection{Problem Statement}
\label{problemStatement:problemStatement}

We define the \emph{reviewer coverage problem} (RCP) as an extension of the \emph{reviewer assignment problem} (RAP) for scientific manuscripts.  
Both problems have the underlying goal of finding suitable \emph{sets of reviewers} for each manuscript. These sets need to be constructed such that (i) reviewer expertise is sufficient for the topics of the respective manuscript, (ii) there are no conflicts of interests between authors of submissions and reviewers, and (iii) overall reviewer load constraints are met.
Contrasting with RAP, RCP does not assume that the current PC is perfect (i.e., has sufficient coverage), but explicitly allows for its extension by adding reviewer candidates from an \emph{extended reviewer candidate pool} (ERC).
So the immediate goal for RCP is the suggestion of new PC members, which leads to sufficient reviewer expertise for all submissions, while also ensuring diversity in the PC in terms of (i) seniority, (ii) location, and (iii) industrial/academic affiliation.\footnote{Gender would also be a desirable diversity aspect for PCs~\cite{SBES}, but we consciously refrain from touching this subject due to the challenges involved in collecting potentially personal information from reviewers for inclusion in our datasets.}
%as automatic dataset generation for this deeply personal characteristic of reviewers is problematic.} 
An additional desirable condition for the inclusion of new PC members is their ability to review multiple papers.
%$RCP$ is not a \textit{typical} item recommendation problem\footnote{We could cast it as one if we would define characteristics of all submissions as traits of a user while the final PC including the newly recommended candidates represents the recommended item.
%All different combinations of reviewers from PC and those from the $ERC$ would correspond to the item space. Most of these items would be excluded from the recommendation step as only those reviewer sets which already contain all PC members would be viable options. However, such a formulation would lead to combinatorical explosion i.e., the set of items would be intractable.}.
%
Formally, the output of RCP is twofold: (1) a ranked list of reviewer suggestions to include in the PC and (2) an assignment of reviewer sets to submissions.

\subsection{Notation}

$M$ describes the set of submissions to a conference for which reviewers from the program committee $PC$ need to be assigned. A single reviewer is addressed as $r_{i}, i \in \{0, \dots, |PC|-1\}$ or only by their index $i$. We address a single submission as $m_{j}, j \in \{0, \dots, |M|-1\}$ or only by their index $j$.
An assignment is \emph{feasible} if all submissions are assigned a predefined number of reviewers $\lambda$, the number of submissions a reviewer is assigned lies between a predefined lower ($\mu^l_i$) and upper bound ($\mu^u_i$), which is specific for each reviewer $i$, and conflicts of interests (COI) are not violated by the assignment. %$\mathcal{A}$ denotes the whole set of feasible assignments $A$:
%
%\begin{align*}
%    \mathcal{A} := \Big\{& A \in \{0, 1\}^{|PC| \times |M|} | \sum_{i \in PC} A_{ij} = \lambda \forall j \in M,\\ & \mu^l_i \le \sum_{j \in M} A_{ij} \le \mu^u_i \forall i \in PC\Big\}
%    \end{align*}
%
The reviewer set assigned to a submission $j$ under a feasible assignment $A$ is denoted by $R_{A}(j)$.
We store similarities of reviewers and submissions in $S \in [0, 1]^{|PC| \times |M|}$; the similarity $S_{ij}$ of reviewer $i$ with submission $j$ is seen as a proxy for expected review quality~\cite{DBLP:journals/corr/abs-1806-06237} \textcolor{black}{and can be determined, e.g., by the cosine similarity between TF-IDF representations of $j$'s and $i$'s profiles, composed of their papers}. In case of a COI between $i$ and $j$, we set $S_{ij} = -1$. We store dependencies between reviewers in $dep \in \{0, 1\}^{|PC| \times |PC|}$; dependencies such as recent co-authorships between reviewers $i$ and $k$ are expressed by $dep_{ik} = 1$ if there is a dependency and $0$ otherwise.

%% file: 4.Method.tex
\section{Method}
\label{method}

%We argue that the diversity aspects we wish for in reviewer assignment need to be an integral part of an approach and thus need to be modelled directly into an algorithm instead of only applying them post-hoc.

\begin{figure}
%\scriptsize
\centering
  \begin{adjustbox}{width=0.8\textwidth}

\begin{tikzpicture}[baseline=(current bounding box.center)]
\node[draw, circle,scale=0.5,fill=gray] (source) at (0,0) {};

\node (s) at (0,2.9) {source};
\node (s) at (0,3.3) {\textbf{L1}};

\draw[dotted] (.75,-2) -- (0.75, 3);

% reviewers
\node[draw, circle,scale=0.5] (r1) at (1.5,0.5) {};
\node[draw, circle,scale=0.5] (r2) at (1.5,0.25) {};
\node[draw, circle,scale=0.5] (r3) at (1.5,-0.25) {};

\node (s) at (1.5,2.925) {reviewer};
\node (s) at (1.5,3.3) {\textbf{L2}};

\draw[dotted] (2.25,-2) -- (2.25, 3);

% decision
\node[draw, circle,scale=0.5] (d1) at (3,1) {};
\node[draw, circle,scale=0.5] (d2) at (3,0.5) {};
\node[draw, circle,scale=0.5] (d3) at (3,0.25) {};
\node[draw, circle,scale=0.5] (d4) at (3,-0.25) {};

\node (s) at (3,2.925) {decision};
\node (s) at (3,3.3) {\textbf{L3}};

\draw[dotted] (3.75,-2) -- (3.75, 3);

% background
\node[draw, circle,scale=0.5] (b1) at (4.5,2.45) {$a_0$};
\node[draw, circle,scale=0.5] (b2) at (4.5,2.1) {$a_1$};
\node[draw, circle,scale=0.5] (b3) at (4.5,1.75) {$a_2$};

\node[draw, circle,scale=0.5] (b4) at (4.5,1.3) {$l_0$};
\node[draw, circle,scale=0.5] (b5) at (4.5,0.6) {$l_6$};

\node[draw, circle,scale=0.5] (b6) at (4.5,0.25) {$l_{0'}$};
\node[draw, circle,scale=0.5] (b7) at (4.5,-0.45) {$l_{6'}$};

\node[draw, circle,scale=0.5] (b8) at (4.5,-0.9) {$s_0$};
\node[draw, circle,scale=0.5] (b9) at (4.5,-1.25) {$s_1$};
\node[draw, circle,scale=0.5] (b10) at (4.5,-1.6) {$s_2$};

\node (s) at (4.5,2.89) {diversity};
\node (s) at (4.5,3.3) {\textbf{L4}};

\draw[dotted] (5.25,-2) -- (5.25, 3);

% paper
\node[draw, circle,scale=0.5] (p1) at (6,0.25) {};
\node[draw, circle,scale=0.5] (p2) at (6,-0.25) {};

\node (s) at (6,2.865) {paper};
\node (s) at (6,3.3) {\textbf{L5}};

\draw[dotted] (6.75,-2) -- (6.75, 3);

% sink
\node[draw, circle,scale=0.5,fill=gray] (sink) at (7.5,0) {} ;

\node (s) at (7.5,2.925) {sink};
\node (s) at (7.5,3.3) {\textbf{L6}};

% dots
\node (dots) at (1.5,-0.) {$\dots$};
\node (dots) at (3,0) {$\dots$};
\node (dots) at (3,0.75) {$\dots$};
\node (dots) at (3,-0.5) {$\dots$};

\node (dots) at (4.5,0.95) {$\dots$};
\node (dots) at (4.5,-0.1) {$\dots$};
\node (dots) at (4.5,-1.95) {$\dots$};

\node (dots) at (6,0) {$\dots$};

% edges
\path[->,draw,above]
(source) edge node {} (r1)
(source) edge node {} (r2)
(source) edge node {} (r3);

\path[->,draw,above]
(r1) edge node {} (d1)
(r1) edge node {} (d2)
(r2) edge node {} (d3)
(r2) edge node {} (d4);

\path[->,draw,above]
(d1) edge node {} (b1)
(d1) edge node {} (b2)
(d1) edge node {} (b3)
(d1) edge node {} (b4)
(d1) edge node {} (b5)
(d1) edge node {} (b6)
(d1) edge node {} (b7)
(d1) edge node {} (b8)
(d1) edge node {} (b9)
(d1) edge node {} (b10)

(d3) edge node {} (b1)
(d3) edge node {} (b2)
(d3) edge node {} (b3)
(d3) edge node {} (b4)
(d3) edge node {} (b5)
(d3) edge node {} (b6)
(d3) edge node {} (b7)
(d3) edge node {} (b8)
(d3) edge node {} (b9)
(d3) edge node {} (b10);

\path[->,draw,above]
(b1) edge node {} (p1)
(b2) edge node {} (p1)
(b3) edge node {} (p1)
(b4) edge node {} (p1)
(b5) edge node {} (p1)
(b6) edge node {} (p1)
(b7) edge node {} (p1)
(b8) edge node {} (p1)
(b9) edge node {} (p1)
(b10) edge node {} (p1);

\path[->,draw,above]
(p1) edge node {} (sink)
(p2) edge node {} (sink);

% bubbles around node types
\draw (4.5,2.1) node[thick,ellipse,dotted,minimum height=1.1cm,minimum width=0.75cm,draw] {};
\draw (4.5,0.425) node[thick,ellipse,dashed,minimum height=2.2cm,minimum width=0.75cm,draw] {};
\draw (4.5,-1.25) node[thick,ellipse,densely dotted,minimum height=1.1cm,minimum width=0.75cm,draw] {};

\end{tikzpicture}
\end{adjustbox}\hspace{5mm}

\begin{tabular}{l"l|l"l|l"l|l}
 node type  &\multicolumn{2}{c"}{Flow via node} &\multicolumn{2}{c"}{I per edge} &\multicolumn{2}{c}{O per edge}\\
            & lb & ub & lb & ub & lb & ub\\ \hline
            
            \hline
    source & 3$*\lambda*$|M| & 3$*\lambda*$|M| & - & - & 3$*l$ & 3$*a$ \\ \hline
    reviewer & 3$*l$ & 3$*a$& 3$*l$ & 3$*a$ & 0 & 3 \\ \hline
    decision & 0 & 3& 0 & 3 & $t$ & $t$\\ \hline
    $a_0$, $a_1$, $s_1$, $s_2$ & 0 & $\lambda$-1 & 0 & 1 & 0 & $\lambda$-1\\
    $a_2$ & 0 & $\lambda$ & 0 & 1 & 1 & $\lambda$\\
    $s_0$ & 1 & $\lambda$ & 0 & 1 & 1 & $\lambda$\\
    $l_0$, \dots, $l_6$ & 0 & ($\lambda$-1)/7 & 0 & 1/7 & 0 & ($\lambda$-1)/7 \\
    $l_{0'}$, \dots, $l_{6'}$ & 1/7 & $\lambda$/7 & 0 & 1/7 & 1/7 & $\lambda$/7\\ \hline
    paper & 3$*\lambda$ & 3$*\lambda$ & $t$& $t$& 3$*\lambda$& 3$*\lambda$ \\ \hline
    sink &  3$*\lambda*$|M| & 3$*\lambda*$|M| & 3$*\lambda$ & 3$*\lambda$ & - & - \\
\end{tabular}

    \caption{\textbf{Top:} A simplified version of the flow network constructed by DiveRS. Only the depicted edges between neighbouring layers allow flow. Background nodes in the dotted ellipse are used to ensure diversity in the professional background, those in the dashed ellipse are used to enforce diversity in the continent of the assigned reviewers and those in the densely dotted ellipse guarantee diversity in seniority.
    \\
    \textbf{Bottom:} Lower (lb) and upper bounds (ub) of incoming (I) and outgoing flow (O) per edge as well as the general flow via a specific node type with ability $a$, demand $\lambda$, lowest load $l$ and amount of flow depending on the node type $t$.}
    \label{fig:flownetwork}
\end{figure}

We introduce \textbf{DiveRS}, a \textbf{Dive}rse \textbf{R}eviewer \textbf{S}uggestion system for extending conference program committees. It focuses not only on fairness of reviewer assignments but also considers diversity in professional background, location of reviewer candidates and their seniority.
We build on and extend a previous state-of-the-art flow-based approach~\cite{DBLP:journals/corr/abs-1806-06237}, by explicitly modelling diversity as a layer in the flow-graph; see Fig.~\ref{fig:flownetwork}.

%\todo{FROM INTRO: Currently we do not treat the PC extension part of RCP as a learning problem due to data sparsity. The PC of previous conferences is mostly passed on to further instances without much additional information except for maybe the explicit exclusion of few candidates. Bidding information of PC members for submissions is not publicly visible in general. Thus, user interactions and ratings are not available for model training purposes.}

%\todo{FROM INTRO: Our reviewing model can be described as follows: each submission is assigned a set of reviewers without conflicts of interest which are most similar to the submission and satisfy the following constraints: (1) Reviewers in the set are independent from each other and (2) contain at least one reviewer from a non-academia (industry) and one from an academia background; (3) Their location (in form of continent) cannot be all the same and (4) at least one senior researcher needs to be in every reviewer set; (5) Reviewers' individual upper and lower bounds of numbers of papers they can review are respected.}

\subsection{Modelling Diversity}

We focus on diversity in three different areas: professional background, location and seniority. We integrate these properties of the assignment in a specific layer in our flow network between papers and reviewers (diversity layer \textbf{L4} in Fig.~\ref{fig:flownetwork}).
Diversity in \emph{professional background} means that each reviewer set has to contain at least one reviewer working in academia and one reviewer (possibly the same one) working in industry.
For diversity in \emph{location} it would be desirable to include reviewers in a reviewer set with locations from completely different geographical locations. The goal here is to not have all reviewers in a set being located on the same continent.
We achieve diversity in \emph{seniority} by enforcing each reviewer set to contain at least one senior researcher~\cite{toronto,tang}. %Reviewers can either be junior, advanced or senior.
\textcolor{black}{Meanwhile, overburdening of reviewers from underrepresented backgrounds can be prevented by decreasing their possible reviewing load. Satisfying all diversity constraints might lead to an increase of the PC size.}

\iffalse
Diversity in \emph{professional background} means that each reviewer set has to contain at least one reviewer working in academia and one reviewer (possibly the same one) working in industry. For each paper, we introduce three nodes symbolising the different professional background types (\emph{industry}, \emph{academia} and \emph{both}). %The upper bound of flow for the first two of these nodes is set to $\kappa -1$, the latter receives a bound of $\kappa$.
For diversity in \emph{location} it would be desirable to include reviewers in a reviewer set with locations from completely different geographical locations. The goal here is to not have all reviewers in a set being located on the same continent. We enforce this property with specific node types in the the diversity layer of our flow network (see nodes $l_x$ in Figure~\ref{fig:flownetwork}).
We achieve diversity in \emph{seniority} by enforcing each reviewer set to contain at least one senior researcher~\cite{toronto,tang}. Reviewers can either be junior, advanced or senior (see nodes $s_x$ in Figure~\ref{fig:flownetwork}).
\fi
%, \textcolor{red}{it would be desirable to include at least one junior, one senior and if the number of reviewers per set is $\ge$ 3 at least one advanced researcher}.

\subsection{Algorithm}

DiveRS identifies submissions with high probability of not obtaining enough suitable (topically fitting and diverse from each other) reviewers and adds new reviewers to the PC accordingly. It then constructs suitable reviewer sets for all submissions from the extended PC. 
%\todo{[High-level description of (main) algorithm: identifying papers for which no suitable (topically matching and diverse) reviewer set can be found.}
Our reviewer suggestion approach is inspired by PR4All~\cite{DBLP:journals/corr/abs-1806-06237}, the current state-of-the-art in flow-based reviewer assignment~\cite{DBLP:conf/kdd/KobrenSM19}.  However, PR4All tackles the reviewer assignment problem (RAP), which is only one element of the larger reviewer coverage problem (RCP) that we are addressing (cf. \S\ref{problemStatement:problemStatement}). 
% roadmap for DiveRS
%Our approach is influenced by PR4All, but as the main difference instead of only focusing on RAP we also tackle RCP. Reviewer assignment is part of ensuring reviewer coverage, 
We do not only construct suitable assignments but also identify possibly problematic papers and actively extend the PC to ensure diverse reviewer sets.

We first discuss the limitations of PR4All in~\S\ref{sec:method:algo:pr4all}, followed by the introduction of DiveRS's reviewer assignment subroutine in~\S\ref{sec:method:algo:DiveRSsubroutine} and its main routine in \S\ref{sec:method:algo:DiveRSmainroutine} which is responsible for identifying problematic submissions and suitable reviewer candidates.
%A heuristic algorithm such as FairFlow~\cite{DBLP:conf/kdd/KobrenSM19} would not satisfy our demand for fairness guarantees.

\subsubsection{PR4All}
\label{sec:method:algo:pr4all}
The goal of PR4All~\cite{DBLP:journals/corr/abs-1806-06237} is the fair assignment of suitable reviewer sets for all submissions with a focus on the most disadvantaged ones. The iterative approach fixes one reviewer set for the worst off submission in each iteration. Each iteration constructs partial reviewer sets for all unassigned submissions consisting of the most similar reviewers. This is their central optimisation problem. The partial sets are merged and considered a possible assignment. One assignment resulting in the highest fairness is computed out of several of these possible assignments. From the best overall merged assignment, the worst off paper is finally assigned its reviewers. Fixed (worst off) papers are disregarded in the next iterative assignment and merge steps until all papers are fixed.

Due to the merge step, PR4All cannot introduce new conditions for the \textcolor{black}{single reviewers and reviewer sets on the final level only}, e.g., lower bounds ($\mu^{l}$) for the number of assigned submissions for each reviewer or that each set must contain at least one reviewer from industry and one from academia. \textcolor{black}{Instead, these lower bounds for reviewers and conditions for reviewer sets would be applied during all parts of the assignment process.}
Overcoming this issue is non-trivial as all partial assignments which are then merged fulfilling the new conditions could \textcolor{black}{also} lead to violated upper bounds ($\mu^{u}$) and an excess of industry reviewers per final \textcolor{black}{reviewer} set. For \textcolor{black}{their initial run with} sets of size 1, the one reviewer would be required to represent both professional backgrounds which is hard to find.
Conditions that only merged assignments have to fulfil cannot be realised in the described optimisation problem.
So, PR4All prevents definition of desirable properties for final assignments that surpass mere similarity, such as diversity in certain properties.

\subsubsection{DiveRS Subroutine: Reviewer Assignment}
\label{sec:method:algo:DiveRSsubroutine}

%Our approach is influenced by PR4All, but strives 
We strive to overcome some of the weaknesses in reviewer assignment encountered in PR4All: we introduce individual upper ($\mu^u_i$) and lower bounds ($\mu^l_i$) of reviewing abilities for each reviewer $i$~\cite{DBLP:conf/kdd/KobrenSM19}. The lower bound describes the number of submissions, a reviewer has to review at least. Additionally, we allow for the definition of dependencies between reviewers \textcolor{black}{(e.g., in case of shared current affiliations or recent collaborations)}, as reviewers in a set should have distinct affiliations to make sure that their opinions are sufficiently independent from each other~\cite{DBLP:conf/nips/JecmenZLSCF20}. The resulting constraint can mathematically be described by the expression $con_{I} : \sum_{j \in M} \sum_{i, k \in R_{A}(j), i \neq k} dep_{ik} == 0$.

Our goal is to assign reviewers to the best fitting submissions to maximise the overall similarity between assigned reviewers and submissions. The following equation formulates the optimisation objective: $maximise J^{S}_{f}(A) := \sum_{i \in R_{A}(j), j \in M} f(S_{ij})$ while all submissions receive $\lambda$ reviewers, dependencies between reviewers, COIs, diversity constraints of reviewer sets as well as reviewers' lower and upper abilities are not violated. $f$ is a monotonically increasing function used to transform similarity values $[0:1]\to[0:\infty]$~\cite{DBLP:journals/corr/abs-1806-06237}. 
%As COI we consider co-authorships of authors of submissions and reviewers in the previous five years and joined current affiliations.

\setlength{\textfloatsep}{10pt} % reduce space below algorithm environment
\begin{algorithm}
\caption{DiveRS main routine: reviewer suggestion for PC extension.}
\label{algo:main_routine}

\begin{algorithmic}[1]
\REQUIRE $\lambda$, $M$, $PC$, $S$, $\mu^l$, $\mu^u$, $dep$, acaInd, location, seniority, $tries$, $S\_ERC$, $\mu^l\_ERC$,  $\mu^u\_ERC$, $dep\_ERC$, acaInd\_ERC, location\_ERC, seniority\_ERC, $\theta$, $\kappa$
\ENSURE Reviewer assignment $A$, problematic papers $M_{<\theta}$

\STATE while PC is not able to produce assignment based on ability and seniority or professional background and |ERC| > 0: include new reviewers from underrepresented aspects with highest average similarities to all manuscripts
\STATE if abilities of PC are not enough to find assignment: terminate with error
%(min(|\{$r_i$\}|: acaInd[$r_i$] == 0, |\{$r_i$\}|: acaInd[$r_i$] == 1) + |\{$r_i$\}|: acaInd[$r_i$] == 2) * ability of respective $r_i$ < $v[\lambda]$ * |$v[M]$|: terminate program with error
\STATE $\forall$ reviewer-submission pairs from $S$ and $S_{ERC}$ set similarity = -1 if similarity < $\theta$  (equivalent to COI)
\STATE $M_{\theta}$ = $M$ w/o submissions with all similarities $<$ $\theta$
\STATE delete reviewers $r$ from $PC$ where $\mu^l_r$ > number of submissions with which they have similarity $\ge$ $\theta$
\STATE $pairs$ = compute all pairs of reviewers in $PC$ and papers in $M_{\theta}$
\WHILE{$sub$($\lambda$, $M_{\theta}$, $PC$, $S$, $[0]^{|PC|}$, $\mu^u$, $dep$, acaInd, location, seniority, $pairs$) does not produce assignment}
\STATE $fewCandidatePapers$ = papers with < $\lambda$ reviewers w/o COI
\STATE run $sub$ multiple times w/o $fewCandidatePapers$ and w/o predefined \% of submissions to identify (possibly problematic) submissions where run fails, i.e., for which no assignment can be computed due to ill-fitting or few reviewers in the submission's area; adjust $pairs$ and $M_{\theta}$ for runs, papers with highest probability of failed run are $problemPapers$
\STATE insert up to $\kappa$ reviewers in $PC$ from ERC fitting $fewCandidatePapers$ + most problematic $problemPapers$ and underrepresented background variables best
\STATE delete papers from $M$ as out of scope for which not enough reviewer ($<$ $\lambda$) candidates with similarity $\ge$ $\theta$ can be found
\STATE $pairs$ = compute all pairs of reviewers and papers
\ENDWHILE
\STATE $f_A$ = [] // list of all feasible assignments 
\FOR{try = 0, try $\le$ $tries$, try ++}
\STATE $pairs_c$ = drop predefined percentage of $pairs$
\STATE $A_{curr}$ = $sub$($\lambda$, $M_{\theta}$, $PC$, $S$, $\mu^l$, $\mu^u$, $dep$, acaInd, location, seniority, $pairs_c$)
\STATE if $A_{curr}$ $\neq$ $\emptyset$ : $f_A$.append($A_{curr}$)
\ENDFOR
\STATE return most diverse assignment from $f_A$, $M$ - $M_{\theta}$
\end{algorithmic}
\end{algorithm}

\begin{algorithm}
\caption{DiveRS subroutine: reviewer assignment step $sub$.}
\label{algo:subroutine}

\begin{algorithmic}[1]
\REQUIRE $\lambda$, $M$, $PC$, $S$, $\mu^l$, $\mu^u$, $dep$, acaInd, location, seniority, pairs, $\theta$
\ENSURE Computed reviewer assignment $A_{curr}$, $\emptyset$ if unfeasible
\STATE \textbf{Initialization:} %set $S_{ij}$ = -1 if < $\theta$, 
flow network (see Figure~\ref{fig:flownetwork}): \\
\textbf{L1} (source, 1 vertex)\\
\textbf{L2} (reviewers, vertex $\forall i \in PC$)\\
%, directed edges from \textbf{L1} to \textbf{L2}, 3 * $\mu^l_i]$ $\le$ $flow[i]$ $\le$ 3 * $\mu^u_i$), \\
\textbf{L3} (reviewer paper decision, vertex $\forall j \in M * \forall i \in PC$)\\
%, $0 \le flow[i * |M| + j]$), \\
\textbf{L4} (diversity, 3 vertex types, vertex $\forall x \in M * 20$, see Figure~\ref{fig:flownetwork})\\
%, $flow[x]$ depends on vertex type, see Figure~\ref{fig:flownetwork}), \\
\textbf{L5} (papers, vertex $\forall j \in M$)\\
%, $flow[j] = 3 * \lambda$), \\
\textbf{L6} (sink, 1 vertex)
%, directed edges from papers to sink, $flow[\omega] = 3 * \lambda * |M|$)

\STATE Reset flow constraints for all vertices in the network: source, reviewer, decision, diversity, papers, sink
\STATE $\forall (ij) \in pairs$: insert edge $(i, j)$ between \textbf{L2} and \textbf{L3} (i.e., set $capacity[i, i * |M| + j] = 3$), adjust flow constraints
\STATE Compute max flow, create assignment $A_{curr}$ corresponding to max flow $\forall (ij)$: \textbf{if} flow on edge $(i, i * |M| + j)$ between \textbf{L2} and \textbf{L3} \textbf{then} assign reviewer $i$ to submission $j$
\STATE return $A_{curr}$
\end{algorithmic}
\end{algorithm}

Algorithm~\ref{algo:main_routine} (main routine) and Algorithm~\ref{algo:subroutine} (subroutine) depict the pseudo code of our approach. %, Figure~\ref{fig:flownetwork} visualises the constructed flow network and associated flow constraints. 
In the subroutine we construct our flow network such that reviewers review a number of submissions limited by their upper and lower bounds. Submissions are reviewed by $\lambda$ reviewers. Each reviewer set for a submission is diverse in professional background (at least one from industry and one from academia), location (not all from the same continent) and seniority (at least one senior reviewer). We only allow the allocation of reviewers to submissions, if this combination is contained in $pairs$. The decision if a reviewer is assigned to a submission is contained in \textbf{L3}, if there is flow over an edge $(i,i*|M| + j)$ between \textbf{L2} (reviewer $i$) and \textbf{L3} (decision to review submission $j$), $i$ is assigned as reviewer for $j$. If we can compute a max flow, we find an feasible assignment.

\subsubsection{DiveRS Main Routine: Reviewer Suggestion for PC Extension}
\label{sec:method:algo:DiveRSmainroutine}

In the main routine we generally first check if the original PC contains enough reviewers such that each submission can be assigned someone from both professional backgrounds as well as one senior reviewer. Otherwise, we include new reviewers with missing diversity properties in the PC from $ERC$ (l. 1). \textcolor{black}{The $ERC$ could, e.g., be composed of authors of former instances of a conference.} 
The similarity threshold $\theta$ heavily influences DiveRS, it defines the minimal similarity between submissions and assigned reviewers~\cite{DBLP:conf/kdd/KobrenSM19} (l. 3-5, 11). If $\theta$=0, the algorithm often finds a solution to the reviewer assignment problem, computed by the subroutine after including new reviewers based on underrepresented diversity aspects (l. 7), and does not need to identify possibly problematic papers (l. 8-9). Those problematic papers (l. 9) are submissions which have a high probability of not getting assigned reviewers (i.e., where runs of the sub-routine oftentimes fail if they are part of $M_{\theta}$).
The higher the value of $\theta$, the more difficult it is to find a feasible assignment.
A value $\kappa$ describes the number of fitting inserted reviewers per iteration (l. 10).
If a feasible assignment (l. 7) has been found for a reviewer set, we randomly exclude reviewers from reviewing submissions in order to find the most diverse assignment (l. 15-20).

Figure~\ref{fig:flownetwork} depicts our network and associated flow constraints. Nodes $a_x$ indicate the professional background of a reviewer ($a_0$ = industry, $a_1$ = academia, $a_2$ = both). The different $l_x$ indicate the location background of reviewers ($l_y$ indicates the presence of a continent in continents associated with a specific reviewer while $l_{y'}$ indicates the continent's absence in their continents; $l_0$ = South America, $l_1$ = Africa, $l_2$ = Antarctica, $l_3$ = Asia, $l_4$ = Oceania, $l_5$ = North America, $l_6$ = Europe). Nodes $s_x$ indicate the different levels of seniority of researchers ($s_0$ = senior, $s_1$ = advanced, $s_2$ = junior).

In our implementation%\footnote{The fully documented code for our approach will be made available online on GitHub at acceptance of the paper.}
, we utilise Gurobi\footnote{\url{https://www.gurobi.com/}}, a commonly used~\cite{DBLP:journals/corr/abs-1806-06237,DBLP:conf/kdd/KobrenSM19} solver software for mathematical optimisation.

%In general, PCs contain only few Asian reviewers compared to their number of submissions~\cite{Severin2021.01.14.426539}.

\subsection{Practical Issues and Effects of Parameters}

Our approach tackles several practical issues which arise in PC extension and reviewer assignment: 

\begin{itemize}
    \item Submissions for which no suitable reviewers can be found as their topics might be out of scope of a current conference can be identified and considered manually.
    \item Reviewers that are part of the original PC should all be assigned at least one submission out of courtesy, even if they might no longer fit the topical composition of the conference. For subsequent instances of the conference, these individuals may no longer be invited to the PC. Our approach identifies such reviewers and is able to assign them to current submissions nevertheless.
\end{itemize}
Running time constraints influence the choice of parameters:
\begin{itemize}
    \item The higher the similarity threshold $\theta$ is set, the more iterations (l. 7-12 A.~\ref{algo:main_routine}) are required until a feasible assignment is found. The higher the number of included new reviewers per run $\kappa$ is set (l. 10 A.~\ref{algo:main_routine}), the longer one single run of the assignment step (A.~\ref{algo:subroutine}) takes but in total less iterations might be needed. If $\kappa$ is high, the total review load will be distributed among the many new candidates. In order to keep the PC comparably small, we advise to have a low $\kappa$ and more iterations in total.
    \item The higher the bias towards incorporation of reviewers with underrepresented background variables (l. 10 A.~\ref{algo:main_routine}), the less focus is put on similarity of reviewers and submissions. In consequence, fairness of assignments decreases while diversity increases. 
\end{itemize}
Note that we do not separately handle sub- or meta-reviewing but DiveRS can be used in these steps with different parametrisation.

%% file: 5.ExperimentalSetup.tex
\section{Experimental Setup}
We present our experimental setup, introducing two new datasets (\S\ref{sec:experimentalSetup:datasets}), our parameter settings (\S\ref{sec:experimentalSetup:generalSetup}), an overview of established measures (\S\ref{sec:experimentalSetup:establishedMeasures}), and novel ones for reviewer assignment assessment, namely diversity and dependency (\S\ref{sec:experimentalSetup:diversity}).

\subsection{Datasets}
\label{sec:experimentalSetup:datasets}

We evaluate on two real-world conference datasets based on the International Conference on the Theory of Information Retrieval (ICTIR) in 2019 (I'19) and 2020 (I'20). 
The data was made available to us by the conference organisers upon request and signing an NDA.
%They are private-use only.
The datasets include all manuscripts submitted to the conferences, not only the accepted ones, authors of submissions, reviewers, real reviewer-submission assignments, and constructed extended reviewer candidate pools.
%We do not distinguish between paper types as program committees are oftentimes published in a merged form.
%
%Table~\ref{tab:ds_overview} contains information on sizes of contained data. 
I'19 (I'20) contains 78 (65) papers submitted by 201 (184) authors. 43 (30) papers were accepted and 36 (35) rejected. There were 43 (67) reviewers. The extended reviewer candidate pool consists of 6,445 (5,692)
%
%The extended reviewer candidate pool %for ECIR'21, I'19 and I'20 
%consists of 
authors from papers which appeared in CIKM, ECIR, ICTIR and SIGIR in the previous five instances of the conferences.
%For CIKM'21 it consists of authors who have at least three papers in the previous five years in the conferences AAAI, ACL, CIKM, CLEF, DASFAA, ECIR, EMNLP, ICDE, ICML, IJCAI, KDD, SIGIR, TREC, WWW as well as the journals ISCI, PVLDB, TKDE and VLDB.

% general information
%For the conferences we have data concerning all submissions, current PC as well as an extended reviewer candidate pool. \textit,{Submission data} contains paper title, abstracts, keywords, acceptance information, affiliations of authors and the authors of papers with COIs. 
%For each author, the datasets contain their DBLP~\cite{ley} key for unique identification and a list of reviewers with COIs. 
For all \textit{reviewers}, we retrieved their DBLP key~\cite{ley}, COIs \textcolor{black}{and dependencies} (collaborators from the previous five years and persons with current shared affiliations), seniority, location, current %and former 
affiliation(s) as well as information on whether they are working in industry and/or academia. 
\textcolor{black}{Demographics were automatically derived from their affiliations and earliest published paper.}
Additionally, we collected the titles of their publications up until the year of the conference, and abstracts from the previous five years for papers which appeared with Springer or ACM.
We performed further manual post-processing to ensure high data quality. % and to resolve any errors. 

\subsection{Parameter Settings} % General Setup
\label{sec:experimentalSetup:generalSetup}

%We evaluate DiveRS on the two introduced datasets, %the publicly available MIDL'20 and ESWC'21 datasets as well as the private
%I'19 and I'20. 

We assign each submission to three reviewers, following the practice of the I'19 and I'20 conferences. % as this reviewer set size was utilised in the actual real assignments.
Similarity between submissions and reviewers (a concatenation of their publications' titles and abstracts) is taken to be the cosine similarity of TF-IDF-weighted document representations, thus all similarity values lie in [0,1]. We utilise $f(S_{ij}) = \frac{1}{1-S_{ij}}$ if $S_{ij} < 1$ and $1*e^{6}$ otherwise \cite{DBLP:journals/corr/abs-1806-06237}.
For DiveRS we set $tries$ = 25, $\kappa$ = 10 and $\theta>0$ to $.25$ for I'20 and $.15$ for I'19.\footnote{Different values for $\theta$ had to be chosen to find feasible assignments, as $\theta$ is highly dependent on topical fit between the submissions and the PC.}
We set $\mu^u$ = 9 for I'19 and = 7 for I'20 according to the real number of maximal assigned submissions per reviewer candidate.
%CIKM'21: 8, ECIR'21: , I'19: 9, I'20: 7.

For the manual evaluations we obtain reliable human assessments by asking respective PC chairs (3 from I'19 and 2 from I'20) to fill out a questionnaire.

\subsection{Established Measures}
\label{sec:experimentalSetup:establishedMeasures}

\iffalse
In this first part of the reviewer assignment evaluation we conduct an automatic evaluation to report the following measures to describe the \textit{quality of the PC}: \textit{Average closeness centrality} and \textit{average betweenness centrality}~\cite{DBLP:conf/jcdl/ZhuangELG07}. Both measures work on the co-author graph of the PC members. The average closeness centrality determines how close PC members are to all others. High closeness indicates fast access to and distribution of new information in the graph and should be found in reputable conferences~\cite{DBLP:conf/jcdl/ZhuangELG07}:
    
    \[CC(PC) = avg_{v \in PC} \left(\frac{n_v - 1}{\sum_{w \in G} shortestPath(v, w)}\right)\]

with $G$ representing the whole unweighted coauthor graph of the scientific community and $n_v$ size of the connected component in which $v$ lies. We construct $G$ by using the DBLP~\cite{ley} dataset.
    
The average betweenness centrality describes the influence of single PC members for the information flow~\cite{DBLP:conf/jcdl/ZhuangELG07}:
    
    \[BC(PC) = avg_{v \in PC} \left(\sum_{w, x \in G} \frac{shortestPath(w, x, via v)}{shortestPath(w, x)}\right)\]
\fi

The following established measures describe the \textit{quality} of reviewer assignments: mean \textit{number of papers} assigned to single reviewers~\cite{DBLP:conf/kdd/KobrenSM19}, \textit{fairness} of the assignment~\cite{DBLP:journals/corr/abs-1806-06237,DBLP:conf/kdd/KobrenSM19}, and \textit{average textual diversity} of reviewer sets~\cite{robustModel}.

\textit{Fairness} of an assignment $A$ is defined as the minimal summed similarity between any submission $j$ and its reviewers $R_{A}(j)$~\cite{DBLP:journals/corr/abs-1806-06237}:
    $ \Gamma^{S}_{f}(A) = min_{j \in M} \left(\sum_{i \in R_{A}(j)} f(S_{ij})\right)  ~,
    $
with $f$ being a monotonically increasing function $[0, 1] \rightarrow [0, \infty]$.
\textit{Average textual diversity} of reviewer sets is calculated by the average Kullback-Leibler (KL) divergence between pairs of reviewers assigned to the submissions~\cite{robustModel}:
    $KL(A) = \\avg_{j \in M} \left(\frac{\sum_{i, k \in R_{A}(j), i\neq k}KL\_Divergence (i, k)}{|R_{A}(j)| * (|R_{A}(j)| - 1)/2}\right) ~.$
We calculate this value on an unigram language model of the reviewer's publication information. Higher values for average KL-divergence indicate less similar reviewers in reviewer sets. Desirable complementary reviewers~\cite{toronto} produce a high value.

\subsection{Novel Measures}
\label{sec:experimentalSetup:diversity}

We present a novel measure for quantifying the \emph{diversity} of backgrounds of reviewers.  
%According to Eq.~\eqref{diversity_eq}, 
We define diversity for reviewers that are part of a feasible assignment $A$, as a linear combination of backgro\-und-, location-, and seniority-based diversity scores (each in $[0,1]$):
$
Div(A) = \textrm{avg}_{j \in M} \left( Div_{BG}(j) + Div_{L}(j) + Div_{S}(j) \right).
$ %\label{diversity_eq}
%\end{equation}
%
Diversity can take values in $[0, 3]$, where higher values are more desirable.
Note that diversity of one single reviewer set $R_A(j)$ can be computed using the same formula by setting $M = \{j\}$.

The component-level diversity scores are estimated as: 
\[Div_{BG}(j) =  1 - \frac{|\sum_{i \in R_A(j)} profBG[i]|}{\lambda}\]
\[Div_{L}(j) = 1 - \frac{1}{{{2} \choose {\lambda}}} * \sum_{i, k \in R_A(j), i \neq k} \frac{|location[i] \cap location[k]|}{|location[i] \cup location[k]|}\]
\[Div_{S}(j) = \sum_{val \in \{0,1,2\}} \mathbb{1}(\exists i \in R_A(j): seniority[i] == val)*\frac{1}{3} ~, \]
where for each reviewer $i$, $profBG[i]$ indicates the professional background (0 if both, -1 if industry, 1 if academia), $location[i]$ denotes the distinct locations associated with $i$, and $seniority[i]$ describing the seniority level (0 if senior, 1 if advanced, 2 if junior). 

We further quantify the \emph{dependency} of an assignment as the percentage of reviewer sets with violated dependencies between reviewers $i$, $k$:
$Dep(A) = \\\frac{\sum_{j \in M} \mathbb{1}(\exists i, k \in R_A(j): i \neq k,  dep_{ik} == 1)}{|M|} * 100$.

\textcolor{black}{\textbf{Example.} Given: $M=\{j\}$, $R_A(j)=\{i$ (both, senior),  $k$ (academia, senior)$\}$, $i$ and $k$ from different locations, $dep_{ik}=0$. We can then compute $Div(A)=(1-\frac{1}{2}) + (1 - \frac{1}{1}*0) + (\frac{1}{3}) = \frac{11}{6}$ and $Dep(A)=0$.}

%% file: 6.Experiments.tex
\section{Experiments}
\label{sec:experiments}

Recall that the output of RCP is twofold: (1) an assignment of reviewer sets to submissions and (2) a ranked list of reviewer suggestions to include in the PC.  We thus divide our evaluation into two parts: an examination of reviewer assignments in~\S\ref{sec:experiments:p1}, using both automatic (\S\ref{sec:experiments:p1:automaticEvaluation}) and manual evaluation (\S\ref{sec:experiments:p1:manualEvaluation}), 
followed by an evaluation of reviewer suggestions using human assessments by the respective PC chairs in~\S\ref{sec:experiments:p2}.

\subsection{Part 1: Reviewer Assignment}
\label{sec:experiments:p1}

For evaluating the reviewer set construction properties of our approach (conducted by our subroutine in \S\ref{sec:method:algo:DiveRSsubroutine}) we compare different variants of our DiveRS (D$_{\theta}$) algorithm against (1) assignments produced by a state-of-the-art flow-based reviewer assignment system, PR4ALL~\cite{DBLP:journals/corr/abs-1806-06237}, and (2) the real reviewer assignments. 

\subsubsection{Automatic Evaluation}
\label{sec:experiments:p1:automaticEvaluation}

In our automatic evaluation, we report the established measures for reviewer assignment from ~\S\ref{sec:experimentalSetup:establishedMeasures}, the newly introduced measures from ~\S\ref{sec:experimentalSetup:diversity}, and the number of unused reviewers from the original PC.

In addition to the DiveRS default setting, we also report on a \emph{restrictive setting}, where each reviewer $i$ from the original PC who can review at least one submission (i.e., similarity $\ge$ $\theta$) needs to be used in the final assignment ($\mu^l_i$ = 1). 
This setting is desirable to prevent displeasing reviewers who have already been invited to the PC by not assigning them to a submission.
In PR4All such an option is not given, including a lower bound for numbers of assignments is impossible as the approach merges assignment sets.

\begin{table}[t]
    \centering
    \caption{Reviewer assignment results for the automatic evaluation in terms of mean workload per reviewer (mW/R) and all initial PC members (/PC), number of unused initial PC members (U) as well as dependency ($Dep$), fairness  ($\Gamma^{S}_{f}$), average textual diversity ($KL$), and diversity ($Div$) of assignments per dataset and method.
    Methods marked with $^*$ correspond to the \emph{restrictive} setting. 
%Mean average number of assignments per reviewer, wPC gives the same value with observation of all initial PC members, the number of unused initial PC members $U$, percentage of sets containing dependencies among reviewers Dep, fairness of assignments $\Gamma^{S}_{f}$, average textual diversity of assignment $KL$ as well as diversity $Div$ per dataset (ds) and method.
    }
    \label{tab:experiments:p1:automaticEvaluation}    
    \begin{tabular}{l|l"l|l|l|l|l|l}
         method & d.set & mW/R (/PC) & U & $Dep$ & $\Gamma^{S}_{f}$ & $KL$ & $Div$\\ \hline
         
        real&  I'19 & 6.16 (5.44) & 5 & 15.38& 2.12 & .45 & 1.51\\
        PR4All& I'19 & 7.09 (5.44) & 10 & 48.72& 3.51 & .52 & 1.58\\
        D$_{\theta=0}$&  I'19 &  6.69 (5.09) & 11& 0& 3.31 & .46 & 2.16\\
        D$_{\theta=0^*}$&  I'19 & 5.09 (5.09) & 0 & 0&3.07 & .45 & 2.13\\
        D$_{\theta>0}$&  I'19 & 6.16 (4.78) & 11 & 0&3.68 & .45 & 2.15\\
        D$_{\theta>0^*}$&  I'19 & 4.98 (4.78) & 2 & 0&3.68 & .45 & 2.13\\
        \hline
        real & I'20 & 3.73 (3.12) & 11 & 24.62& 2.4 & .45 & 1.57\\
        PR4All& I'20 & 4.88 (2.91) & 27 & 47.69& 3.62 & .52 & 1.55\\
        D$_{\theta=0}$& I'20 & 4.53 (2.87) & 25 & 0& 3.5 & .47  & 2.04\\
        D$_{\theta=0^*}$& I'20 & 2.87 (2.87) & 0 & 0& 3.18 & .47 & 2.05\\
        D$_{\theta>0}$& I'20 & 3.16 (2.03) & 32 & 0& 4.05 & .44 & 2.12\\
        D$_{\theta>0^*}$& I'20 & 2.23 (2.13) & 4 & 0& 4.05 & .44 & 2.09\\
       
    \end{tabular}
\end{table}

%For I'19 (I'20) 15.38\% (24.62\%) of real assignments and 48.72\% (47.69\%) of those computed by PR4All contain dependencies between reviewers in a set. 

Table~\ref{tab:experiments:p1:automaticEvaluation} reports the results of the automatic evaluation. DiveRS achieves the highest diversity scores regardless of the setting. 
Real assignments are worse in fairness and diversity than the automatically constructed sets. Usage of D$_{\theta>0}$ leads to fairer and mostly more diverse results compared to the D$_{\theta=0}$variants. KL-divergence does not seem to change much between configurations, but PR4All produces sets with the highest score.
Introduction of new PC members naturally reduces the mean workload per reviewer. 
With the restrictive DiveRS variants to include all reviewers from the original PC in the assignment (marked with $^*$), we achieve fairness, KL, and diversity values comparable to the unrestricted variants. For unrestricted DiveRS versions, the number of unused reviewers from the original PC also lies around the value produced by PR4All.
Of all methods, it is only DiveRS that prevents the generation of assignments with dependencies between reviewers in sets.

For I'20 with D$_{\theta=.25}$ we found four papers as well as four original reviewers which were out of scope of the conference. For I'19 with D$_{\theta=.15}$ we found two original reviewers which were out of scope of the conference.

\subsubsection{Manual Evaluation}
\label{sec:experiments:p1:manualEvaluation}

We set up an online questionnaire where the two respective groups of PC chairs assessed the suitability of reviewer sets for ten randomly drawn submissions for their conferences.  We presented them with four reviewer sets:\footnote{If sets produced from different methods are identical, we only depict it once.} the real assignment as well as three automatic assignments produced by PR4All, D$_{\theta=0}$, and D$_{\theta>0}$. % (As before, we set $\theta>0$ to $.15$ for I'19 and to $.25$ for I'20.)  
For each assignment, PC chairs indicated the set's suitability on a four-point scale (\textit{no reviewers are suitable, two reviewers need to be replaced, one reviewer needs to be replaced, suitable assignment}) and justified their decision in a free-text field.

\begin{figure}
	\centering
  \includegraphics[width=0.78\textwidth]{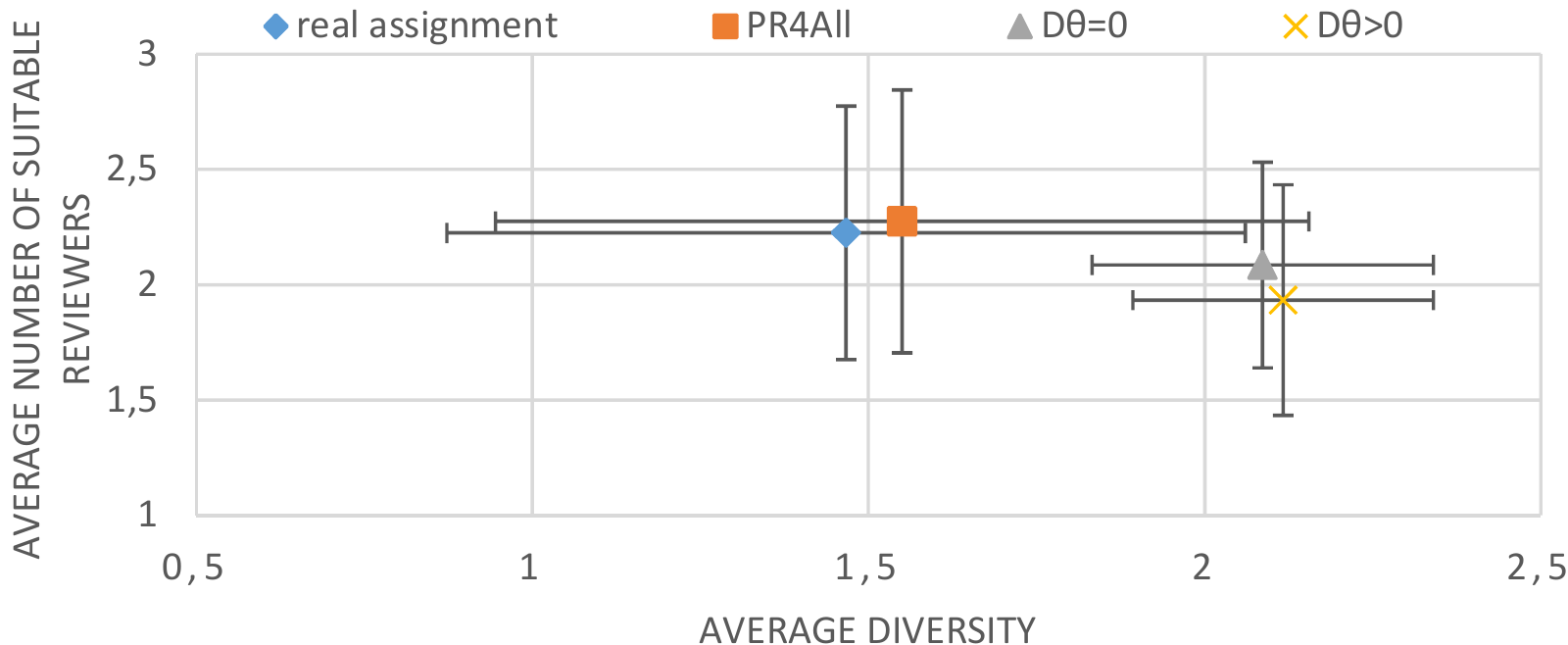}
  
	\caption{Reviewer assignment results using manual evaluation, displaying average diversity (x-axis) against the number of suitable reviewers (y-axis). The error bars correspond to the standard deviation per method.  Results are reported on the two datasets combined.}
	\label{fig:experiments:p1:manualEvaluation}
\end{figure}

Figure~\ref{fig:experiments:p1:manualEvaluation} shows the average diversity against the number of suitable reviewers, for the two datasets combined. Both $D_{\theta=0}$ and $D_{\theta>0}$ produce reviewer sets with fewer suitable reviewers than the real assignment and PR4All---on the other hand, they produce much more diverse assignments. 
We observed low agreement between PC chairs when asked about the suitability of reviewer sets, as reflected in the standard deviations.  It suggests that there are additional factors that may need to be considered in the reviewer assignment task; the free text comments, however, did not allow us to identify any common patterns. 
%
%The agreements of PC chair's rankings computed with Fleiss' Kappa for I'20 (I'19) pt 1 are as follows: \textbf{in general} .0123 (-.0196); \textbf{per method} for PR: .0141 (-.3333), for RA: 0 (0), for DiveRS$_{\theta=0}$: 0 (-.0294), for DiveRS$_{\theta=.25}$: -0.125 (.2157).
%
%Krippendorff's Alpha for inter-rater reliability (dependency between values and number of occurrences of class is assumed): \textbf{in general} -0.1953 (-.0571)
%
%Only few PC chairs justified their decisions such that we were not able to find common explanations for suitability assessments.

\subsubsection{Summary of Findings}

DiveRS achieves fairness values which are comparable to those achieved by PR4All, without specifically focusing on this aspect of the problem. Additionally, our approach introduces more options to control reviewer load and to ensure the independence of reviewers. % in sets to be independent from each other which highly contrasts the number of sets with dependencies constructed by the other methods. 
The resulting diversity values for DiveRS are much better than those of the real assignments or PR4All.

In our experiments, we found that there is a high probability of not assigning papers to all reviewers from the initial PC. Some members might have been included in a PC solely due to their reputation, not because of current interests or expertise in the fields of the submissions~\cite{DBLP:conf/cikm/BordeaBB13}. Unlike other methods, DiveRS offers the possibility of enforcing the involvement of all (fitting) PC members.

Manual evaluation showed the difficulty of objectively assessing the suitability of reviewer assignments, as we observed a high degree of disagreement between PC chairs. A comparison of diversity against the number of suitable reviewers revealed DiveRS' tendency to sacrifice some suitability in order to achieve high diversity.

% most reviewers only have casual knowledge in most manuscripts, reviewing them would be time consuming and maybe unreliable \cite{DBLP:conf/fun/BenderMS0V16}

\subsection{Part 2: Reviewer Suggestion}
\label{sec:experiments:p2}

In the second part of the evaluation, we measure the quality of reviewer suggestions for inclusion in the PC; this corresponds to our main routine (\S\ref{sec:method:algo:DiveRSmainroutine}). 
We consider up to ten reviewer candidates suggested by DiveRS (6 for I'19 and 10 for I'20\footnote{DiveRS introduces different numbers of reviewers based on the dataset as well as $\theta$.}). %We again set $\theta>0$ to $.25$ for I'20 and $.15$ for I'19. 
PC chairs are given a list of reviewers that could be invited. Each candidate is presented by their name, link to their DBLP profile, their main diversity attributes (professional background, location, seniority) as well as an explanation why they would be useful for an exemplary submission (e.g., \textit{non-academia and academia background, topically fitting}). Additionally, other submissions in which the suggested candidate could help are listed.
PC chairs are then asked to rate the relevance of the suggestion, their confidence in their assessment, as well as the usefulness of the explanation and how convincing it is on a Likert scale from 1 (not at all) to 5 (very).

%With the assessment of PC chairs with the newly included reviewers we report nDCG@k for the list of inserted candidates. We also report the agreement of the respective PC chairs. 

\begin{table}[t]
    \caption{Reviewer suggestion results, listing average values for relevance of explanation (r), confidence (f), usefulness (u), convincingness (c), as well as suggestion ranking (NDCG) per dataset.
    Usefulness and convincingness are further subdivided (in parentheses) to cases with relevance below 3 (u$_{<}$, c$_{<}$) and above 3 (u$_{>}$, c$_{>}$).}
    \label{tab:experiments:p2}
    \centering
    \begin{tabular}{l"l|l|l|l|l}
        d.set & r & f & u (u$_{<}$/u$_{>}$) & c (c$_{<}$/c$_{>}$) & NDCG \\ \hline
        I'19 & 2.22 & 4.06 & 2.56 (2.15/3.67) & 2.06 (1.69/3) & .7967 \\
        I'20 & 2.65 & 3.65 & 2.2 (1.89/2.17) & 2.25 (1.56/3.17) & .9105 \\
    \end{tabular}
\end{table}

The PC chairs' agreement on the relevance of suggestions is low for both datasets, which leads us to believe that this task is also very difficult to evaluate.
%
%PC chair's agreement on relevance of suggestions computed with Fleiss' $\kappa$ for I'19 (I'20) is .28 (.0361).
%
%For I'19 (I'20) average values of observed dimensions are as follows: relevance of explanation is 2.22 (2.65), confidence is 4.06 (3.65), usefulness is 2.56 (2.2) and convincingness of the explanation is 2.06 (2.25). In I'19 (I'20) average usefulness of explanations is 3.67 (2.17) and average convincingness 3 (3.17) in cases where relevance of suggestions was rated as 4 or 5. In cases where relevance was rated 1 or 2 average usefulness of explanations is 2.15 (1.89) and average convincingness 1.69 (1.56).
The average values for assessed quality dimensions of suggested reviewer candidates are listed in Table~\ref{tab:experiments:p2}.  In general, relevancy for suggested reviewers is low, usefulness and convincingness of explanations increase drastically if only relevant (relevancy>3) are considered. 
We also evaluate suggestions as a ranked list in terms of NDCG, and observe high scores, especially for I'20. This can be interpreted as our method's ability to estimate the confidence of the recommendations and rank them accordingly.
\textcolor{black}{Our results hint at difficulties in suggestions' quality assessment, which should be investigated further to make findings more conclusive.}

%number of satisfied bids as reviewer assignment evaluation criterion if this information is not used in the assignment process~\cite{DBLP:journals/corr/abs-1806-06237}

%\textcolor{red}{nDCG if graded bidding is available (2/3 wants to review, 1 willing to review, 0 not willing to review)}

%\textcolor{red}{include: plots showing the improvement/change of evaluation measures when adding more batches of recommended reviewers to the PC}

%\textcolor{red}{We also conduct experiments with the accepted and rejected papers to evaluate whether papers could get rejected because of low reviewer coverage.}

%\textcolor{red}{The number of newly recommended members is fixed for batches}

%% file: 7.Conclusion.tex
\section{Conclusion}
In this paper we introduced the novel reviewer coverage problem and proposed DiveRS, a flow-based reviewer assignment and PC member suggestion approach to solve it. DiveRS constructs diverse and fair reviewer set assignments for submissions and also suggests new reviewer candidates for inclusion in the PC. Our evaluation on two real world datasets showed DiveRS' superior diversity compared to both real assignments and the current state-of-the-art.  Our experiments also highlighted the inherent difficulties of the reviewer assignment task, as evidenced by the low inter-annotator agreement between former PC chairs.

% future work
Future work could include utilising bidding information, when available, to identify papers with insufficient coverage.
\textcolor{black}{Requiring junior reviewers to be part of each reviewer set may be desirable at times.}
%  covered with sufficiently similar or interested reviewers as well as help in the assignment process. 
Also, candidate suggestions may be subjected to stricter requirements, e.g., they should be able to review multiple submissions or not be considered at all. \textcolor{black}{Additionally, creating a reusable dataset for reviewer suggestion will be a challenge in itself.}
Finally, there are further gains to be made by employing more advanced methods for determining the similarity between reviewers and submissions. %Here for example squared cosine similarity or document embedding methods could be applied.

%Another line of work could try to simplify or reformulate the constraints from our constructed optimisation problem to allow for application on huge conferences.
%Other attempts could be made to include even more or fine grained diversity aspects of reviewer which would require for open datasets.